\documentclass[final,3p,twocolumn]{elsarticle}
\usepackage{multicol}
\usepackage{amssymb}
\usepackage{amsmath}
\usepackage{graphicx}
\graphicspath{{figures/}{\main/figures/}}
\usepackage{dcolumn}
\usepackage{etoolbox}
\usepackage{siunitx}
\usepackage{float}
\usepackage{caption}
\usepackage{subcaption}
\usepackage{multirow}
\usepackage{rotating}
\usepackage{hyperref}
\usepackage{xcolor}
\hypersetup{
    colorlinks=true,
    linkcolor=blue,
    urlcolor=blue,
    citecolor=blue,
    }

\begin{document}
\begin{frontmatter}
\title{Conical Targets for Enhanced \\High-Current Positron Sources}

%% Author affiliation
\author[PSI,CERN,EPFL]{Nicolas~Vallis}
\ead{nicolas.vallis@cern.ch}
\author[CERN]{Ramiro~Mena-Andrade}
\ead{ramiro.francisco.mena.andrade@cern.ch}
\author[CERN]{Barbara~Humann}
\author[CERN]{Yongke~Zhao}
\author[PSI]{Paolo~Craievich}
\author[CERN]{Jean-Louis~Grenard}
\author[CERN]{Andrea~Latina}
\author[CERN]{Anton~Lechner}
\author[CERN]{Antonio~Perillo-Marcone}
\author[PSI]{Riccardo~Zennaro}
\author[IJCLAB]{Fahad~Alharthi}
\author[IJCLAB]{Iryna~Chaikovska}
\author[IJCLAB]{Robert~Chehab}
\author[PSI,EPFL]{Mike~Seidel}

\affiliation[PSI]{organization={Paul Scherrer Institute},
            city={Villigen},
            country={Switzerland}}
\affiliation[CERN]{organization={CERN},
            city={Geneva},
            country={Switzerland}}
\affiliation[IJCLAB]{organization={CNRS/IN2P3, IJCLab},
            city={Orsay},
            country={France}}
\affiliation[EPFL]{organization={\'{E}cole Polytechnique F\'{e}d\'{e}rale de Lausanne},
            city={Lausanne},
            country={Switzerland}}

%% Abstract
\begin{abstract}
Previous pair-production-driven positron source designs have assumed that the transverse dimension of the target is significantly greater than the secondary beam it generates. This paper explores the use of targets with different transverse profiles with the aim of enhancing positron production. The starting point of this research is the concept of wire targets, proposed by M. James \textit{et al.} in 1991~\cite{James:WireTargets} for the former SLC positron source. Building on this foundation, this study takes this concept a step further by introducing conical-shaped targets, which can substantially improve the yield by reducing the reabsorption of positrons by the target—an issue that is worsened by the high-field solenoid lenses commonly used for positron capture. Using Geant4 simulations, we propose new conical targets adapted for the parameters of the future collider FCC-ee and its positron source test facility P\textsuperscript{3} (PSI Positron Production experiment) at the Paul Scherrer Institute. We find that conical targets can nearly double the positron production at the target and enhance the baseline positron yield of FCC-ee by around 60\%. Additionally, we present the thermo-mechanical studies for the conical targets based on the FCC-ee primary beam power requirements and outline the mechanical implementation for a proof-of-principle demonstration at the future P\textsuperscript{3} facility.
\end{abstract}

\end{frontmatter}

\section{\label{sec:intro}Introduction}

Pair-production-driven positron (e$^{+}$) sources are a proven technique to produce intense e$^{+}$ beams, making them essential for high-current machines like lepton colliders. Typically, this process is initiated by sending a high-energy electron (e$^{-}$) beam into a target made of high-Z material. Upon impact, the e$^{-}$ beam starts an electromagnetic shower leading to a high yield of secondary e$^{+}$e$^{-}$ pairs~\cite{chehab:CAS}. As the showers develop, some secondary e$^{+}$ exit the target and contribute to the yield, while others annihilate with an e$^{-}$~\cite{RPP2022}. Conventionally, targets are assumed to be disk-shaped, with a semi-infinite transverse size compared to the shower's \textit{Moliere radius}~\cite{Bock:Detectors}. Thus, it can be assumed that all e$^{+}$ exit from the downstream surface of the target. The concept of wire targets, introduced by M. James et al.~\cite{James:WireTargets}, proposed reducing the target radius to limit the development of particle showers and enable a prompt exit of e$^{+}$ from the sides of the target, thereby increasing the total e$^{+}$ yield. Originally, this concept was suggested as an upgrade to the former SLC e$^{+}$ source, based on a 33~GeV primary e$^{-}$ beam with a beam size of 0.6~mm and a tungsten target of 6~$X_0$ thickness, $X_0$ being the radiation length. From a practical standpoint, this study concluded that using a wire-shaped target with a 10~$X_0$ length and a 0.5~mm radius could triple the baseline e$^+$ yield.

\begin{figure}[b!]
    \centering
    \begin{subfigure}[b]{0.43\textwidth}
        \includegraphics[width=\textwidth]{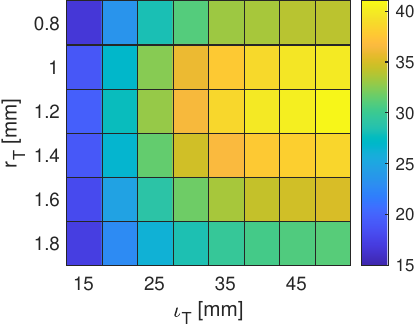}
        \caption{\label{fig:YieldCylnoAMDa}}
    \end{subfigure}
    \begin{subfigure}[b]{0.43\textwidth}
                \includegraphics[width=\textwidth]{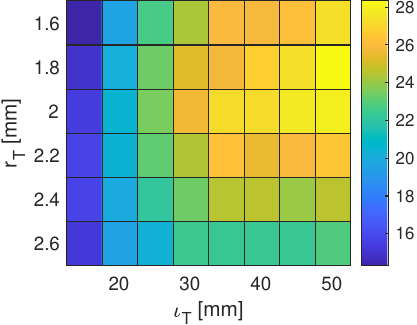}
        \caption{\label{fig:YieldCylnoAMDb}}
    \end{subfigure}
    \caption{\label{fig:YieldCylnoAMD} e$^{+}$ yield without HTS solenoid field for primary $\sigma_x=$~0.5~mm (a) and $\sigma_x=$~1~mm (b). Based on Geant4 simulations with 10\textsuperscript{4} primary events. }
\end{figure}

This paper explores the use of converter targets with a finite transverse size for the e$^{+}$ source of CERN's Future Circular Collider (FCC)~\cite{FCC-ee_CDR, Craievich:FCCeeInjector} and its test facility at the Paul Scherrer Institute, known as P\textsuperscript{3} (or \textit{P-cubed})~\cite{Vallis:P3}. Based on the parameters presented in FCC's midterm report~\cite{FCC-MTR2024}, these e$^+$ sources use a conventional pair-production scheme, driven by a 6~GeV e$^{-}$ beam and a fixed tungsten target. However, a key innovation of these designs is the use of a high-temperature superconducting (HTS) solenoid around the target, enabling the first highly efficient e$^{+}$ capture system thanks to a field strength of 12.7~T. Using Geant4~\cite{G4} and FLUKA~\cite{Fluka2015}, this study adapts the wire targets concept for these future e$^{+}$ sources and expands it into a further optimized conical profile that enhances compatibility with the HTS solenoid. The primary advantage of conical profiles is to minimize e$^{+}$ reabsorption within the target, an issue intensified by the 12.7~T focusing field. The paper includes a thorough thermo-mechanical analysis with FCC-ee's primary beam power parameters as presented the FCC midterm report, consisting of trains of two e$^-$ bunches at 6~GeV, in the nano-Coulomb range, and with a repetition rate of 200~Hz. Finally, we also outline the mechanical implementation of different conical targets in P\textsuperscript{3} for a future proof-of-principle demonstration.

\section{Wire Targets for FCC-ee and P\textsuperscript{3}}
\label{sec:VIIWireTargets}

\begin{figure}[b!]
    \centering
    \begin{subfigure}[b]{0.43\textwidth}
        \includegraphics[width=\textwidth]{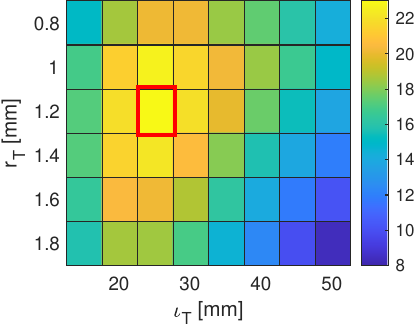}
        \caption{\label{fig:YieldCylAMDa}}
    \end{subfigure}
    \begin{subfigure}[b]{0.43\textwidth}
                \includegraphics[width=\textwidth]{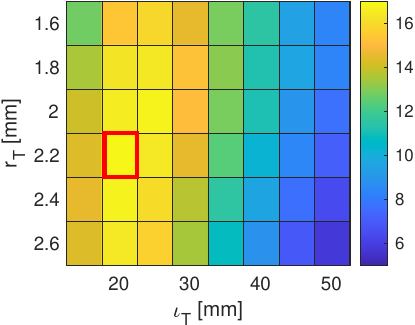}
        \caption{\label{fig:YieldCylAMDb}}
    \end{subfigure}
    \caption{\label{fig:YieldCylAMD} e$^{+}$ yield including the HTS solenoid field for primary $\sigma_x=$~0.5~mm (a) and $\sigma_x=$~1~mm (b). Based on Geant4 simulations with 10\textsuperscript{4} primary events. Maxima in red.}
\end{figure}

\begin{table*}
\caption{Overview of baseline and wire targets for P\textsuperscript{3} and FCC-ee. Relative enhancement with respect to baseline e$^{+}$ yield 13.77 is shown in parenthesis. }
\label{tab:WireTargetOverview}
\centering
\begin{tabular}{lccc}
	 & \textbf{Baseline}  &  \textbf{Wire ($\sigma_x=$~0.5~mm)} &  \textbf{Wire ($\sigma_x=$~1~mm)} \\
        \hline
        $E_0$ [GeV]  & 6 & 6 & 6  \\
        $\sigma_x$  & 1 & 0.5 & 1  \\
        \hline
        $r_T$ [mm] & semi-infinite & 1.2 & 2.2  \\
        $\ell_T$ [mm] & 17.5 (5~$X_0$) & 25 (7.14~$X_0$) & 20 (5.71~$X_0$) \\
        Yield at target w/o HTS & 13.77 & 32.9 (+ 140 \%) & 20.4 (+ 49 \%) \\
        Yield at target w/ HTS & 13.77 & 22.9 (+ 67 \%) & 16.8 (+ 23 \%)\\
        $\sigma_{px}$ [MeV/c] & 7.1 & 7.5 & 7.38 \\
        $\sigma_{E}$ [MeV] & 123 & 176 & 274 \\
        \hline
\end{tabular}
\end{table*}

\begin{figure}[b!]
    \centering
    \begin{subfigure}[b]{0.475\textwidth}
        \includegraphics[width=\textwidth]{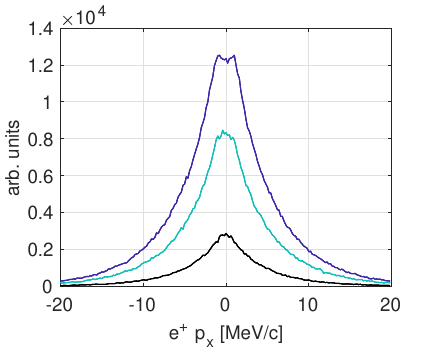}
        \caption{\label{fig:CylPropertiesa}}
    \end{subfigure}
    \begin{subfigure}[b]{0.475\textwidth}
                \includegraphics[width=\textwidth]{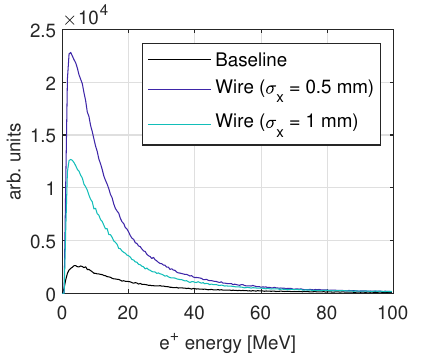}
        \caption{\label{fig:CylPropertiesb}}
    \end{subfigure}
    \caption{\label{fig:CylProperties} Histograms of transverse momentum (a) and energy spectrum (b) of e$^{+}$ distributions generated by baseline and wire targets.}
\end{figure}

Up to now, design studies for FCC-ee and P\textsuperscript{3} have assumed that the transverse dimension of the target is significantly larger than that of the generated secondary beam. Thus, the optimization of the target geometry has been reduced to a scan of its thickness. The baseline optimal thickness is 17.5~mm (equivalent to 5~$X_0$), which provides a yield of 13.77 e$^+$ exiting the target per primary e$^{-}$~\cite{Chaikosvka:FCCee23}. Using Geant4 simulations, we revised this assumption and studied the impact of the transverse size of the target on the yield. The first step of this study is to adapt the concept of wire targets to our design, and find the target radius ($r_t$) and thickness ($\ell_t$) that maximize the e$^{+}$ yield. In this paper, the e$^{+}$ yield is defined as the number of e$^{+}$ that reach the downstream plane of the target per primary e$^{-}$. Two options are considered for the primary e$^{-}$ beam size: $\sigma_x=$~0.5~mm and $\sigma_x=$~1~mm. 

Fig.~\ref{fig:YieldCylnoAMD} shows the results of the conducted parameter scans, without including the HTS solenoid focusing field. While there is a clear $r_t$ optimum, the yield increases with the length of the target. While this yield gain can be of almost a factor of~4, a more realistic assessment is shown in Fig.~\ref{fig:YieldCylAMD}, which includes the effect of the HTS solenoid. It is observed that the yield gain is significantly moderated, as the target reabsorbs many e$^{+}$ due to the spiraling motion induced by the 12.7~T magnetic field. However, the yield enhancement is still significant. The optimum working points are found at ($r_t$,~$\ell_t$)|$_{\sigma_x=0.5}$~=~(1.2,~25)~mm and ($r_t$,~$\ell_t$)|$_{\sigma_x=1}$~~=~(2.2,~20)~mm which would provide e$^{+}$ yields of 22.9 and 16.8 respectively. Compared to the baseline yield of 13.77, the use of a wire target provides an enhancement of +67\% for the $\sigma_x=$~0.5~mm case and +23\% for $\sigma_x=$~1~mm.

\begin{table*}
\caption{Overview of baseline and conical targets for P\textsuperscript{3} and FCC-ee. Relative enhancement with respect to baseline e$^{+}$ yield 13.77 is shown in parenthesis.}
\label{tab:ConicalTargetOverview}
\centering
\begin{tabular}{lccc}
	 & \textbf{Baseline}  &  \textbf{Cone ($\sigma_x=$~0.5~mm)} &  \textbf{Cone ($\sigma_x=$~1~mm)} \\
        \hline
        $E_0$ [GeV]  & 6 & 6 & 6  \\
        $\sigma_x$  & 1 & 0.5 & 1  \\
        \hline
        $r_{up}$ [mm] & semi-infinite & 1.8 & 3.8  \\
        $r_{down}$ [mm]  & semi-infinite & 0.6 & 1.5  \\
        $\ell_T$ [mm] & 17.5 (5~$X_0$) & 25 (7.14~$X_0$) & 21 (6~$X_0$) \\
        Yield at target w/ HTS & 13.77 & 25.89 (+ 89 \%) & 18.92 (+ 37 \%) \\
        Yield at DR & 6.5\textsuperscript{a}& 11.0 (+ 69 \%) & 7.8 (+ 20 \%)\\
        Yield at DR w/ support & 6.5 & 10.56 (+ 62 \%) & 7.56 (+ 16 \%)\\
        \hline
        \multicolumn{4}{l}{\footnotesize\textsuperscript{a} $\sigma_x$~=~0.5~mm would provide yield of 7.}
\end{tabular}
\end{table*}

A key issue of pair-production-driven e$^{+}$ sources is the beam quality, with the typical transverse emittance and energy spread being orders of magnitude higher than the average e$^{-}$ machine. We analyzed whether targets with a finite transverse size enhance these properties, described in Fig.~\ref{fig:CylProperties} and Table~\ref{tab:WireTargetOverview}. The results indicate that the transverse momentum ($\sigma_{px}$) is relatively unaffected, while the energy spread ($\sigma_E$) increases. However, this increase in $\sigma_E$ is attributed to a higher number of high-energy e$^{+}$ emerging from the wire targets. Importantly, as shown in Fig.~\ref{fig:CylProperties}, we cannot conclude that the energy spectrum is significantly degraded, as the population of e$^{+}$ in the multi-MeV range grows substantially, where they have the highest probability of being captured.

\section{Conical Targets}
\label{sec:VIIConicalTargets}

\begin{figure}[b!]
    \centering
    \begin{subfigure}[b]{0.43\textwidth}
        \includegraphics[width=\textwidth]{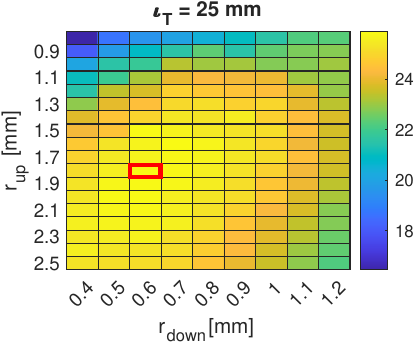}
        \caption{\label{fig:YieldConea}}
    \end{subfigure}
    \begin{subfigure}[b]{0.43\textwidth}
        \includegraphics[width=\textwidth]{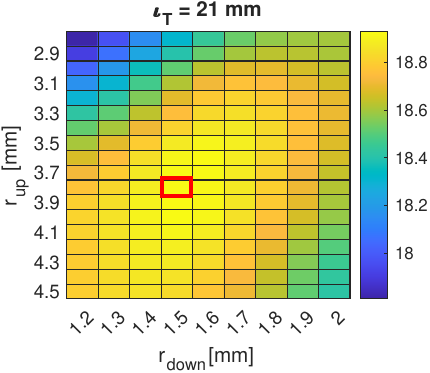}
        \caption{\label{fig:YieldConeb}}
    \end{subfigure}
    \caption{\label{fig:YieldCone} Total e$^{+}$ yield at the conical target's downstream plane with HTS field for the $\sigma_x=$~0.5~mm (a) and $\sigma_x=$~1~mm (b) options. Based on Geant4 simulations with 10\textsuperscript{4} primary events. The optimum working points are marked in red.}
\end{figure}

Conical-shaped targets can substantially improve the performance of wire targets. First, the conical geometry can slightly increase the number of e$^{+}$ that emerge from the target. However, the yield is most significantly enhanced by reducing the number of e$^{+}$ reabsorbed by the target, as the conical profile minimizes the secondary impact probability. The new scheme for the geometry optimization is a truncated cone defined by three parameters: the radii of the cone's upstream ($r_{up}$) and downstream ($r_{down}$) ends and its length ($\ell_T$). In addition, it must be noted that from this point onward, simulations always include the HTS solenoid field. The optimization process involves several iterations. First, the $\ell_T$ values are fixed according to the value previously found in the wire target optimizations of section~\ref{sec:VIIWireTargets}, and $r_{up}$ and $r_{down}$ are swept in pursuit of a new maximum e$^{+}$ yield. Next, $r_{up}$ and $r_{down}$ are fixed and $\ell_T$ is swept again to enhance the previous result. This process is repeated until there is no significant margin for improvement. Fig.~\ref{fig:YieldCone} illustrates the scans of $r_{up}$ and $r_{down}$ during the final iteration, and Table~\ref{tab:ConicalTargetOverview} summarizes the results. The yield maxima are found at ($r_{up}$,~$r_{down}$,~$\ell_T$)|$_{\sigma_x=0.5}$~=~(1.8,~0.6,~25)~mm and ($r_{up}$,~$r_{down}$,~$\ell_T$)|$_{\sigma_x=1}$~=~(3.8,~1.5,~21)~mm. In both cases, they would represent an improvement of +13\% compared to their wire target counterpart, and +89\% and +37\% relative to the baseline design~\cite{Vallis:P3}.

\begin{figure}[t!]
    \centering
    \includegraphics[width=0.5\textwidth]{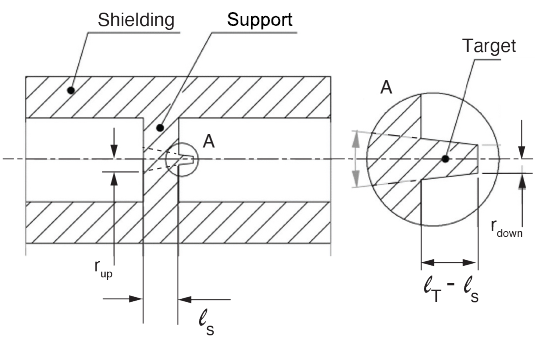} 
    \caption{\label{fig:TargetInterfaceScheme} Longitudinal cross-section of a conical target embedded in a tungsten support.}
\end{figure}

\begin{figure}[b!]
    \centering
    \includegraphics[width=0.5\textwidth]{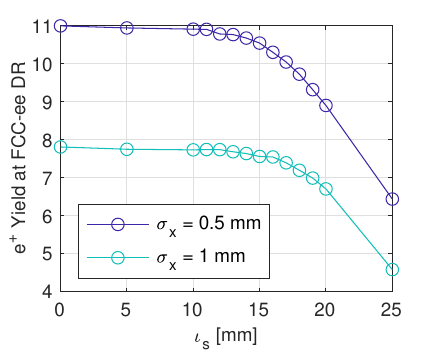} 
    \caption{\label{fig:YieldConewithSupport} Simulated e$^{+}$ yield at the FCC-ee DR for different values of $\ell_s$. Based on Geant4 and RF-Track simulations.} %Courtesy of Y. Zhao (CERN).}
\end{figure}

The simulation framework for the conical target layout was extended to evaluate its impact on the e$^{+}$ yield at the FCC-ee injector damping ring (DR). For this analysis, the optimized conical geometries were incorporated into the comprehensive simulation setup for the FCC-ee e$^{+}$ source and capture linac~\cite{Zhao:AMDOptimization}. This framework includes all major accelerator and beamline components from the target to the DR and combines Geant4 simulations with RF-Track~\cite{RF-Track} to simulate e$^{+}$ production and beam dynamics up to the nominal DR energy of 1.54~GeV. Finally, the yield at the DR is calculated by applying a longitudinal acceptance window of one RF bucket in time and $\pm$3.8$\%$ in energy. The results indicate that the provided e$^{+}$ yield at the DR would be 11.0 in the $\sigma_x=$~0.5~mm case and 7.8 for $\sigma_x=$~1~mm, which represent an increase of 69\% and 20\% respectively relative to the baseline yield of 6.5~\cite{CHART23}. It is important to note that the yield gain in the case $\sigma_x=$~0.5~mm is not only attributed to the conical geometry but also to the beam size reduction. Previous design versions of the FCC-ee e$^{+}$ source employed a primary beam with $\sigma_x=$~0.5~mm, which would raise the baseline DR yield to 7 when using a disk-shaped target. Nevertheless, even with this beam size, adopting a conical target geometry alone would increase the yield at the DR by 57\%.
Despite the significant yield gains offered by conical targets, their implementation requires a physical supporting structure, which could impact performance. As seen in Fig.~\ref{fig:TargetInterfaceScheme}, this support is conceptualized as a tungsten monolith, consisting of a tungsten disk of thickness $\ell_s<\ell_T$ from which the downstream part of the conical target protrudes. Consequently, the effective length of the conical profile is reduced by $\ell_T-\ell_s$, which may have a detrimental effect on the e$^{+}$ yields calculated in Table~\ref{tab:ConicalTargetOverview}. To quantify these losses, the tungsten support was incorporated into the model used for the e$^{+}$ yield simulations. A scan of $\ell_s$ is shown in Fig.~\ref{fig:YieldConewithSupport}, where it is observed that the yield remains nearly unaffected up to around $\ell_s$~=~15~mm for both $\sigma_x$ options. At this working point, the final e$^{+}$ yield at the DR would be 10.56 in the $\sigma_x=$~0.5~mm case and 7.56 for $\sigma_x=$~1~mm, preserving most of the gains achieved by the conical geometry.

%\vfill\null
%\columnbreak
\section{Thermo-Mechanical Aspects}
\label{sec:Thermo-Mechanical_Aspects}

\begin{table*}
\caption{Overview of thermal load at baseline (as presented in FCC's midterm report) and conical targets based on current FCC-ee parameters~\cite{CHART23}.}
\label{tab:ConicalTargetThermal}
\centering
\begin{tabular}{lccc}
	 & \textbf{Baseline}  &  \textbf{Cone ($\sigma_x=$~0.5~mm)} &  \textbf{Cone ($\sigma_x=$~1~mm)} \\
        \hline
        $E_0$ [GeV]  & 6 & 6 & 6  \\
        $\sigma_x$ [mm]  & 1 & 0.5 & 1  \\
        Bunch charge [nC]  &2.08&1.325&1.775\\%& 1.3 & 0.827 & 1.11  \\
        Rep. rate [Hz]   & 200 & 200 & 200 \\
        Bunches per pulse [Hz]   & 2 & 2 & 2 \\
        Beam power [kW]   & 4.99 & 3.18 & 4.26 \\
        Power deposited [kW]   & 1.46 & 1.66 & 1.79 \\
        Max. Power density [kW/cm$^3$] &10.86 &15.83 &9.31\\
        \hline
        $r_{up}$ [mm] & semi-infinite & 1.8 & 3.8  \\
        $r_{down}$ [mm]  & semi-infinite & 0.6 & 1.5  \\
        $\ell_T$ [mm] & 17.5 (5~$X_0$) & 25 (7.14~$X_0$) & 21 (6~$X_0$) \\
        $\ell_s$ [mm] & - & 15 & 15 \\
        \hline
        Max. Temperature [$^{\circ}$C] & 303 & 2201 & 1033 \\
        Max. Equivalent Stress [MPa] & 152 & 349 & 442 \\
        \hline
\end{tabular}
\end{table*}

\begin{figure}[b!]
    \centering
    \includegraphics[width=0.5\textwidth]{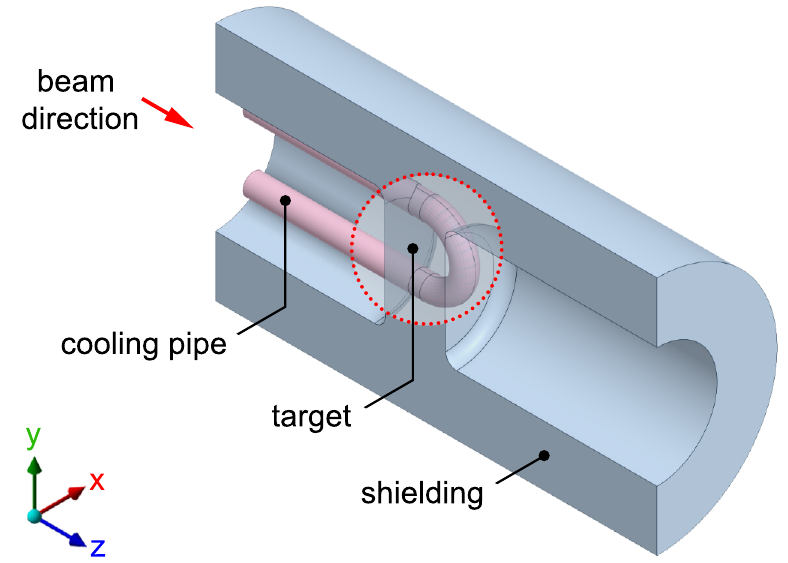} 
    \caption{\label{fig:CoolingConcept} FCC-ee e$^+$ source target baseline design. The detailed zone shows the embedded tantalum cooling pipes. Due to the symmetry respect to the x-axis, only one half of the geometry is shown.}
\end{figure}

The FCC-ee primary e$^{-}$ beam parameters, listed in Table~\ref{tab:ConicalTargetThermal}, are extremely stringent and require a cooling system. For this purpose, a pressurized water cooling circuit is included in the target design for FCC-ee, shown in Fig.~\ref{fig:CoolingConcept}, which consists of a pair of embedded tantalum pipes that transport water from an upstream source and circulate through a 180$^{\circ}$ elbow inside the tungsten core. This setup will allow the beam-impacted region to properly transfer the heat and avoid the direct contact of water with bare tungsten. In contrast, the thermal load on the P\textsuperscript{3} target is not a significant concern due to the low power of the primary e$^{-}$ beam (1.2 W). Thus, the resulting average power deposited on the target is negligible (0.31 W), and no cooling system is needed. However, this cooling system is included on the P\textsuperscript{3} target as a manufacturing R\&D demonstrator. %\pagebreak

%Model setup
This section aims to determine whether the proposed conical targets can withstand the primary e$^{-}$ beam of FCC-ee and be integrated into its current water cooling system. To this end, both conical configurations were analyzed using the commercial Finite Element (FE) software \mbox{ANSYS}~\cite{ANSYS} and compared with the target baseline design. Fig.~\ref{fig:ConicalSimLayouts} depicts the geometrical models used for thermo-mechanical simulations for all three configurations: the baseline and the two conical geometries.

\subsection{Model setup}
\begin{figure*}[ht!]
    \centering
    \includegraphics[width=\textwidth]{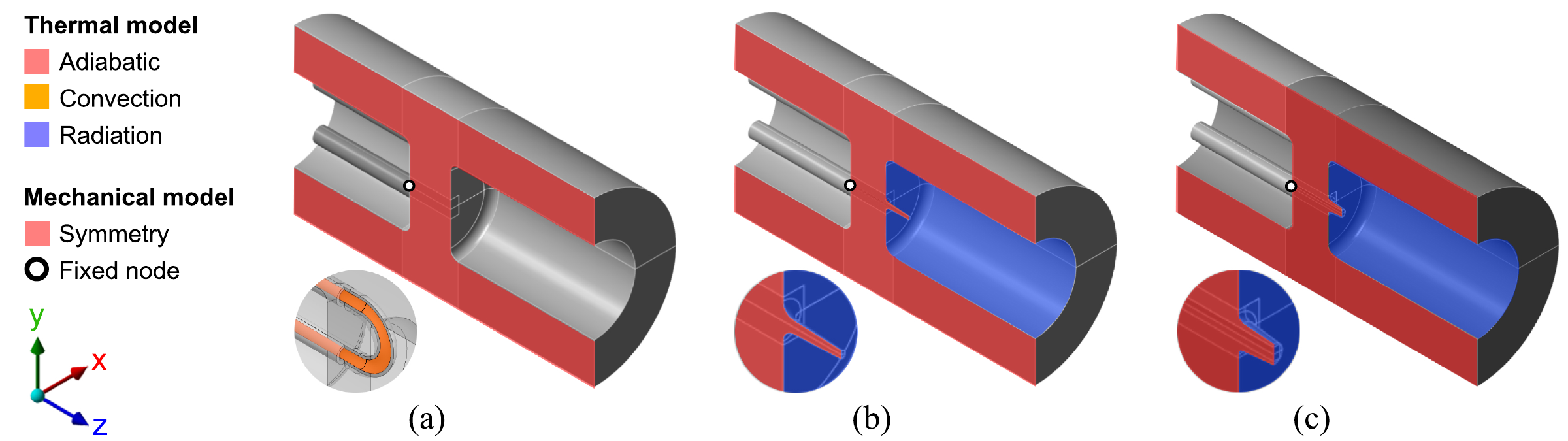} 
    \caption{\label{fig:ConicalSimLayouts} Thermo-mechanical model setup used for the baseline (a), $\sigma_x=$~0.5~mm conical (b), and $\sigma_x=$~1~mm conical (c) targets. Convection along the inner walls of cooling pipes are applied in all models but shown only for case (a).}
\end{figure*}

The simulation models consist of a tungsten monoblock, which integrates the shielding, the support and the protruding conical target in one single part. In both conical targets, the thickness of the support is set at $\ell_s=$~15~mm, which, as noted in section~\ref{sec:VIIConicalTargets}, is the highest possible thickness without a significant yield decrease, while for the baseline case, the target thickness is $\ell_T=$17.5 mm. For the thermal model, the main input is the power deposited by the beam on the target. The generated heat propagates by \emph{(i)} conduction through the solid, \emph{(ii)} convection along the cooling pipes\footnote{The cooling system uses water at 27~$^\circ$C, flowing at a rate of 5~m/s and with a pressure of 20~bar.}, and \emph{(iii)} radiation from the outer surface of the cone into the downstream tungsten walls. As a simplification hypothesis, the involved surfaces subjected to radiation are assumed to be an ideal black body (i.e emissivity $\varepsilon$~=~1). Due to the symmetry  of the model with respect to the x-axis, the surface located in the yz-plane (depicted in red in Fig.~\ref{fig:ConicalSimLayouts}) is considered as adiabatic. Then, for the mechanical model, the previously obtained temperatures are used as an input and due to the presence of a non-homogeneous thermal field, the resulting thermal stresses are calculated. Symmetric boundary conditions are applied along the adiabatic surface and the node located along the beam axis with coordinates \mbox{(x,~y,~z) = (0,~0,~17.5)} mm is fixed to prevent any rigid body motion, as shown in Fig.~\ref{fig:ConicalSimLayouts}.

\begin{figure*}[ht!]
\centering
    \includegraphics[width=0.8\textwidth]{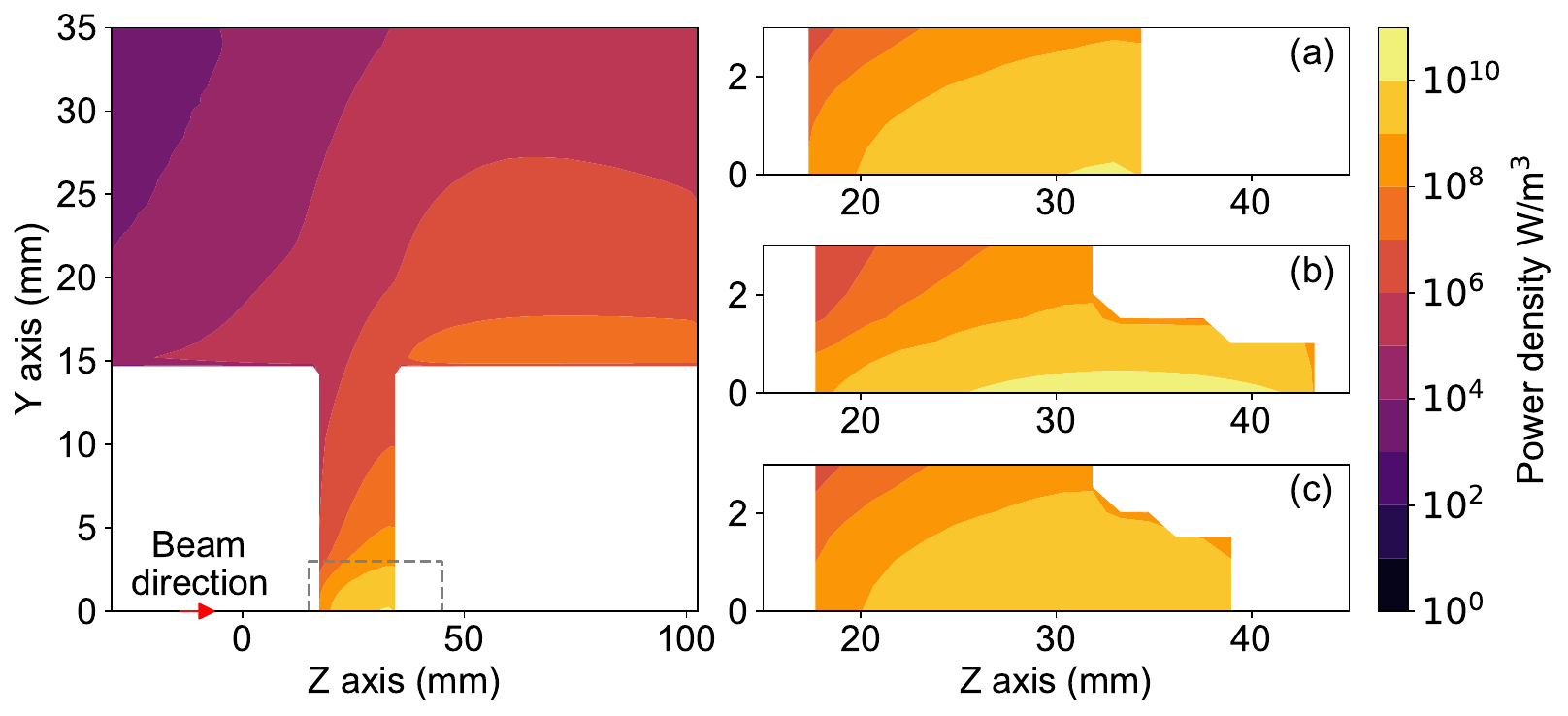} 
    \caption{\label{fig:Power_Density_FLUKA} Power density deposition map obtained from FLUKA used as input for the thermo-mechanical simulations for the baseline (a), $\sigma_x=$~0.5~mm conical (b) and $\sigma_x=$~1~mm conical (c) targets.}
\end{figure*}

\begin{figure*}[t!]
    \centering
    \begin{subfigure}[b]{\textwidth}
        \includegraphics[width=\textwidth]{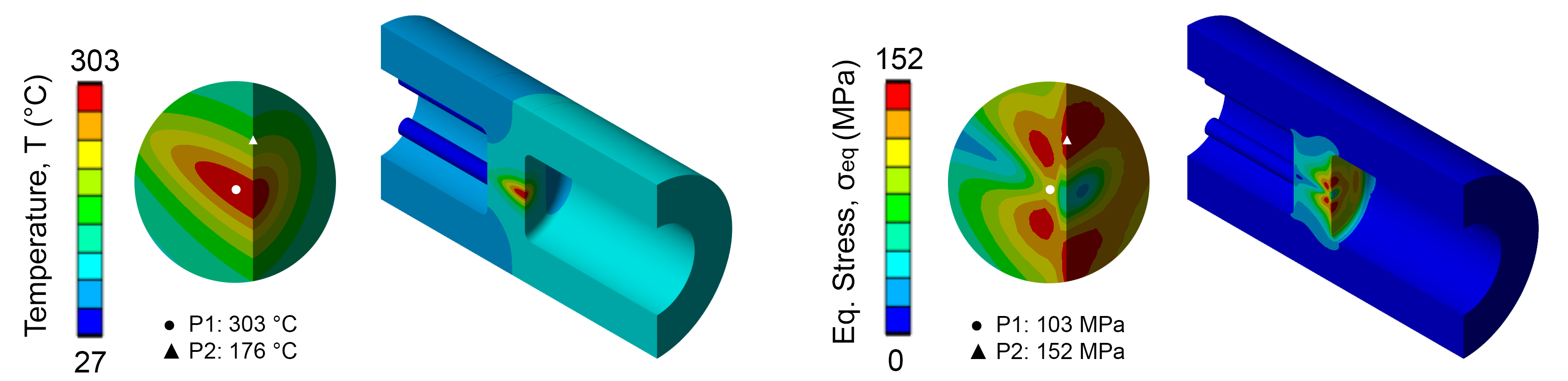}
        \caption{\label{fig:ThermalSimulationa}}
    \end{subfigure}
    \begin{subfigure}[b]{\textwidth}
                \includegraphics[width=\textwidth]{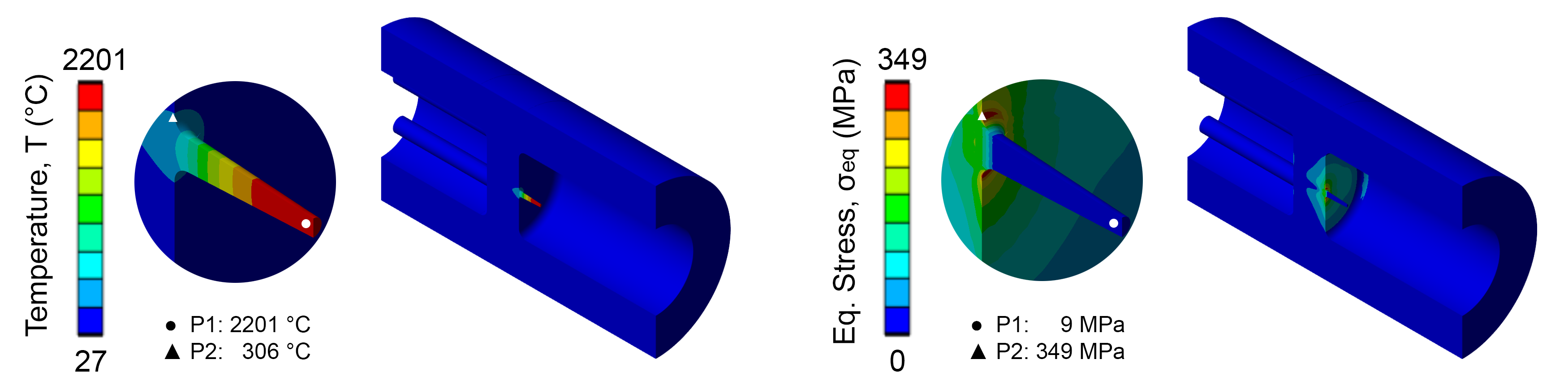}
        \caption{\label{fig:ThermalSimulationb}}
    \end{subfigure}
        \begin{subfigure}[b]{\textwidth}
                \includegraphics[width=\textwidth]{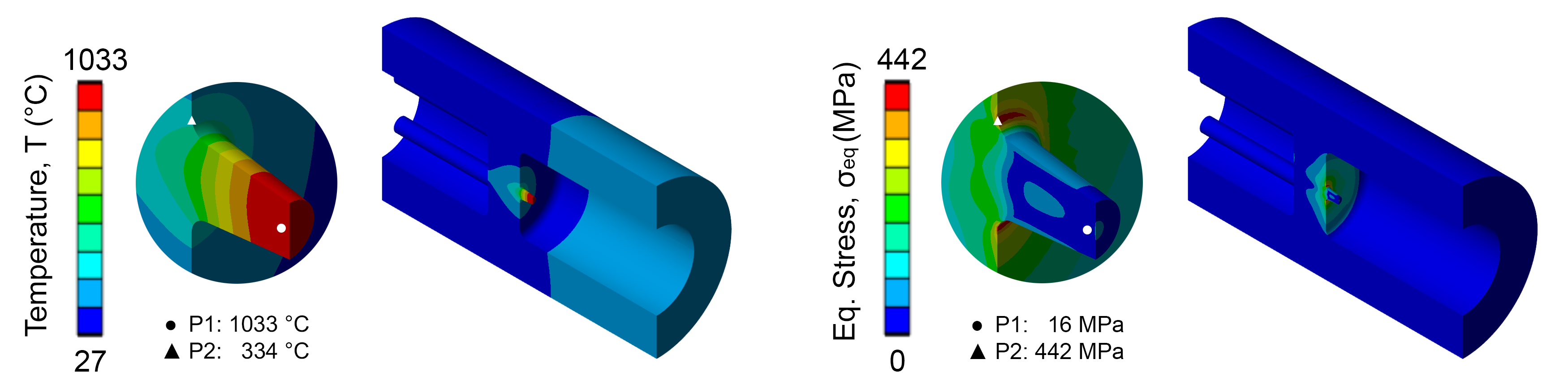}
        \caption{\label{fig:ThermalSimulationc}}
    \end{subfigure}
    \caption{\label{fig:ThermalSimulation} Steady-state thermo-mechanical results: temperature (left) and equivalent stress (right) distributions for the baseline (a), $\sigma_x=$~0.5~mm conical (b) and $\sigma_x=$~1~mm conical (c) targets. Maximum temperature and stress points are marked with a circle and triangle, respectively.}
\end{figure*}

\subsection{Power density deposition}

%Q_0 definition
According to the current FCC-ee design requirements, the injector must provide trains of two e$^{+}$ and two e$^{-}$ bunches into the booster ring, with a single bunch charge of 5~nC and a repetition rate of 200~Hz. On top of that, the FCC-ee injector design collaboration fixed a goal of 5.4~nC e$^{+}$ bunches accepted at the DR, as well as an overall safety margin of 2.5, to compensate for potential unforeseen losses~\cite{CHART23}. Namely, the primary e$^{-}$ bunch charge ($Q_0$) is defined by:
\begin{equation}
\label{eq:e_bunch_charge}
    Q_0 = \frac{5.4\ nC\ \times\ 2.5}{e^+\ Yield\ at\ DR}
\end{equation}
%Impact on power deposition
As seen in table~\ref{tab:ConicalTargetThermal}, the e$^{+}$ yield increase provided by the conical targets allows for a reduction in $Q_0$, resulting in lower beam power. However, according to Geant4 and FLUKA simulations, using conical targets also increases the amount of power deposited in the target. This counter-intuitive result is better understood by analyzing the power density map obtained from the Monte Carlo simulations as shown in Fig.~\ref{fig:Power_Density_FLUKA}. Here the same logarithmic scale is used for all cases, the iso-contour power density lines are plotted and due to the radial symmetry of the FLUKA model, only a 2D plot is shown. While a similar power density is deposited on the shielding for all cases, a zoom in the beam impact region reveals a bigger zone for the peak value  obtained for the $\sigma_x=$~0.5~mm conic target as shown in Fig.~\ref{fig:Power_Density_FLUKA}b. This is caused by combining a smaller beam size $\sigma_x$ with a longer target thickness $\ell_T$, resulting in a beam with a higher energy density and more material volume available to develop the electromagnetic shower when compared to the other configurations.

\subsection{Steady-state analysis}

\begin{figure*}
    \centering
    \begin{subfigure}[b]{0.493\textwidth}
        \includegraphics[width=\textwidth]{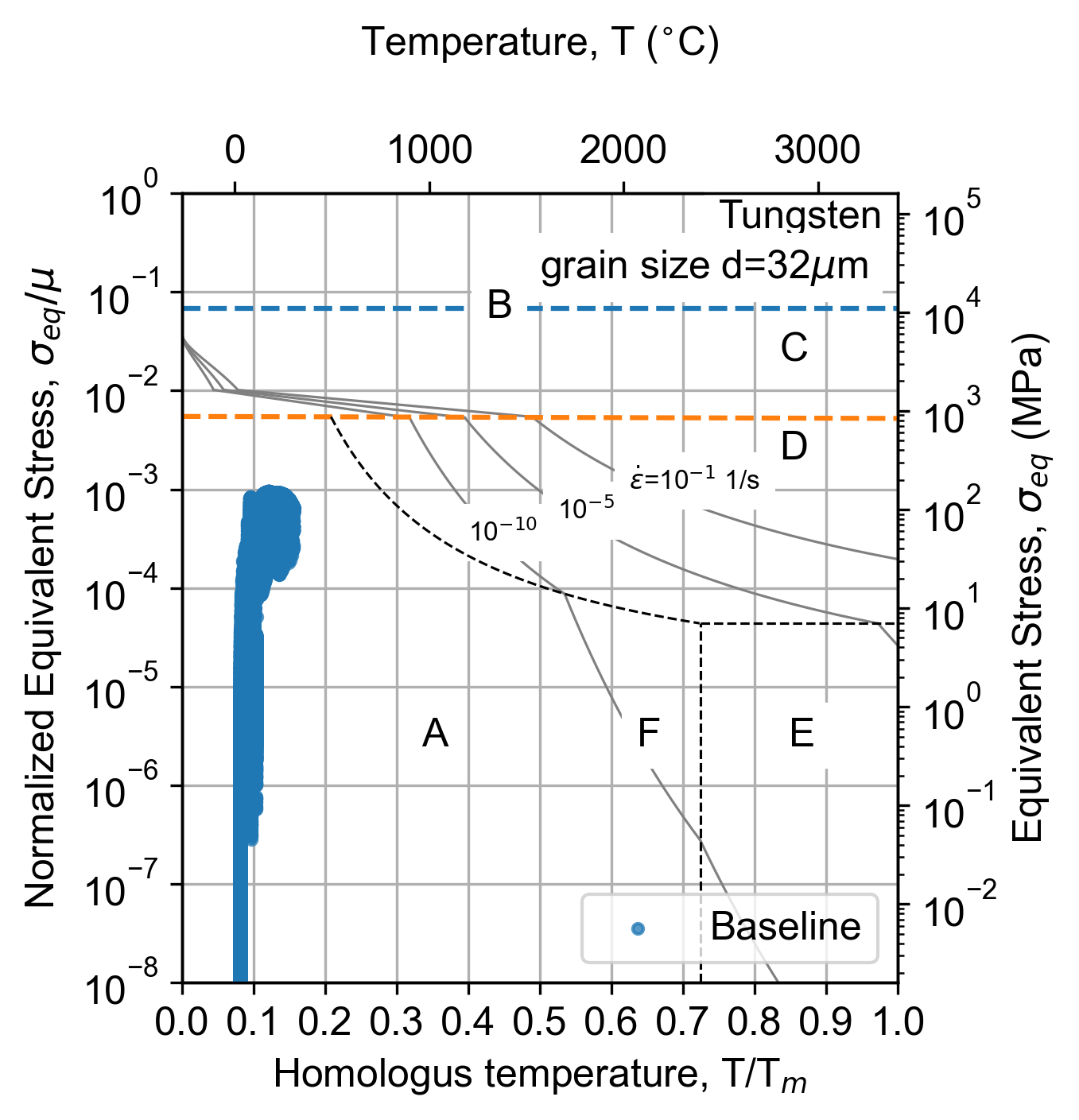}
        \caption{\label{fig:Stress_Temp_baseline}}
    \end{subfigure}
    \begin{subfigure}[b]{0.493\textwidth}
        \includegraphics[width=\textwidth]{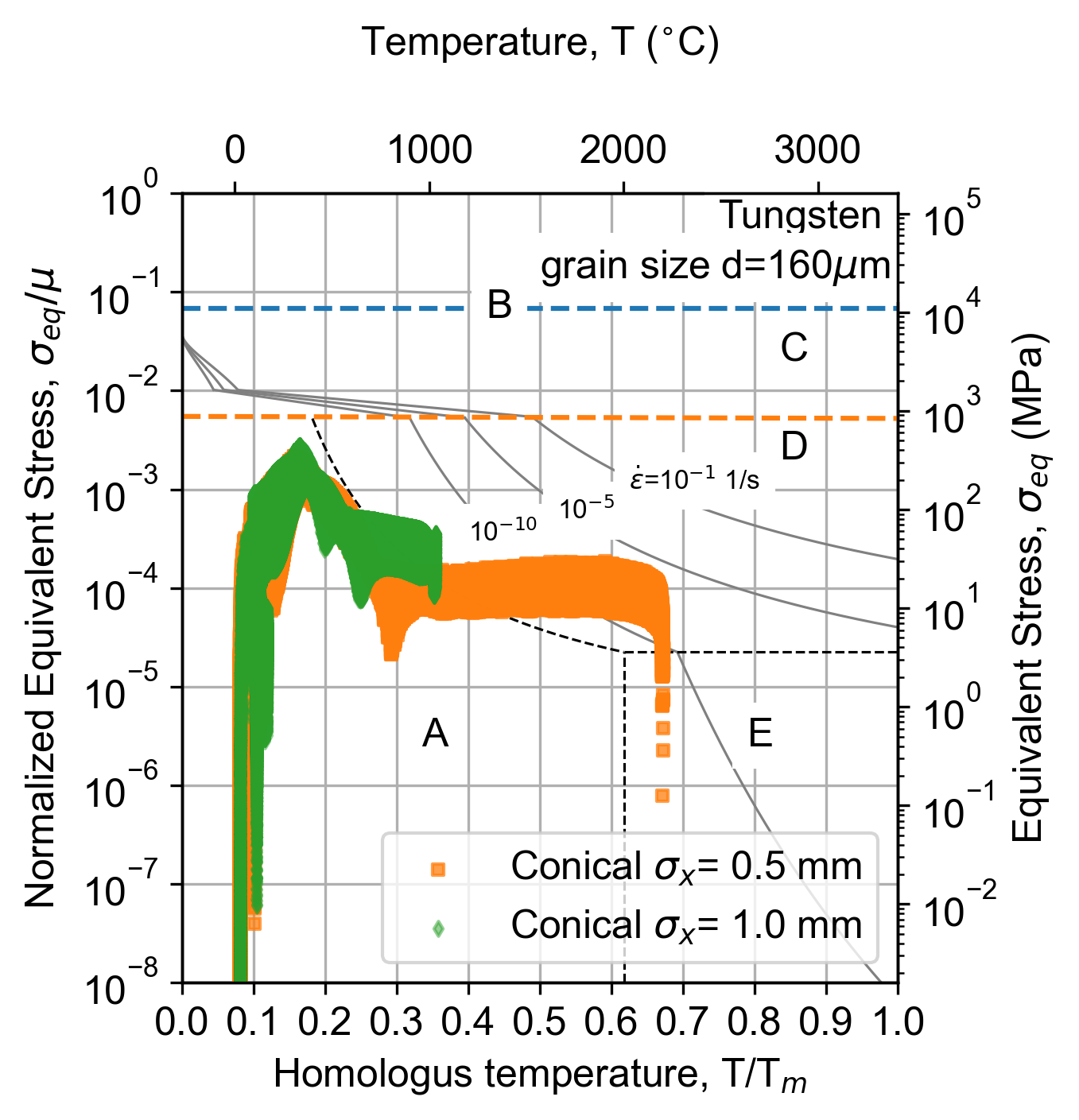}
        \caption{\label{fig:Stress_Temp_Conical}}
    \end{subfigure}
    \caption{\label{fig:Deformation_mech_map} Deformation mechanism maps for polycrystalline tungsten \cite{Ashby1972}: baseline (a) and conical (b) targets. Legend: A: Elastic regime, B: Theoretical shear stress, C: Dislocation glide, D: Dislocation creep, E: Diffusional flow (Nabarro creep) and F: Diffusional flow (Coble creep).}
\end{figure*}

%Steady-state results
Fig.~\ref{fig:ThermalSimulation} shows the steady-state thermo-mechanical results. Note that the location of maximum temperature is not coincident with the position of maximum equivalent thermal stresses. Hereafter, each location is referred to as P1 and P2 respectively. While P1 takes place along the beam axis, close to the exit surface for the baseline and near the protrusion tip's end for both conical targets, P2 is located next to the strong  gradient of temperature (or material), at the exit surface for the baseline design and at the rounded fillet on the support for both conical cases. The maximum temperature values ordered progressively at P1 are 303, 1033 and 2201~$^\circ$C, corresponding to the baseline, $\sigma_x=$~1~mm and $\sigma_x=$~0.5~mm cases, respectively. To classify these results, the absolute melting temperature $T_m$ for  tungsten is used (3695 K), so that, the baseline and $\sigma_x=$~1~mm conical target work in the medium temperature regime $T\in[ 0.1T_m-0.5T_m)$, while the remaining configuration works in the high temperature region $T\in[ 0.5T_m-0.9T_m)$. In terms of stresses at P1, the low values registered for both conical targets (9 and 16 MPa) are due to the homogeneous temperature reached on the conical tip that allows a relatively free expansion of the protrusion, in contrast to the 103 MPa registered on the baseline design that is constrained by the surrounding material of the target. On the other hand, P2 presents lower temperature values but the maximum equivalent stresses (152, 349 and 442 MPa for the baseline, $\sigma_x=$~0.5~mm and $\sigma_x=$~1~mm cases) are all located at the surface level and although the values are below the ultimate stress $\sigma_u$ at its associated temperatures (see values in Table \ref{tab:Material_properties_for_fatigue}), they can be potentially candidates to develop a crack in case of thermal fatigue. 

Given the elevated temperatures reached on both conical targets during the steady-state, recrystallization followed by grain growth can be triggered. For pure metals, recrystallization occurs around 0.4$T_m$ \cite{Callister2015}, and it depends on the material microstructure obtained via prior deformation, and the holding duration time at a certain temperature \cite{Alfonso2014, Alfonso2015}. For pure tungsten, the recrystallization temperature $T_r$ is reported in the interval of 1000-1550 $^{\circ}$C, for severely deformed and cold rolled samples, respectively \cite{Suslova2014}. However, for long term exposition (time $\geq$ 200 hours), hot-rolled pure W recrystallized by annealing at 1100 $^{\circ}$C \cite{Tsuchida2018}. The reduction of mechanical properties of recrystallized tungsten (e.g. lower yield/tensile strength \cite{Wirtz2013}) can affect negatively the performance of both conical targets.
\vfill\null
\columnbreak

%Deformation mechanism maps
Fig.~\ref{fig:Deformation_mech_map} shows the deformation-mechanism maps for polycrystalline tungsten \cite{Ashby1972}. By plotting the fields of equivalent stresses against temperature, it is possible to identify the resulting working regime of the material. In addition, this tool allows us to include the strain rate $\dot{\varepsilon}$ (1/s) and the material grain size $d$ ($\mu$m) to compare the expected material behavior for each configuration. The plot uses the melting temperature $T_m$ (K) and the shear modulus at room temperature $\mu$ (MPa) to normalize the results. For example, for the baseline target, it is assumed a grain size of 32 $\mu$m. Here, all points are located in zone A, corresponding to the elastic regime and its location is not sensitive to the strain rate effects. On the other hand, as recrystallization and grain growth take place for both conical targets, it is assumed a bigger grain size of 160 $\mu$m. Under this configuration, zone D (dislocation creep) and above (i.e zones E and F) can expand or shrink based on a defined strain rate. For example, given a quasi-static range $\dot{\varepsilon} \in [10^{-5},10^{-1}]$ (1/s), both conical targets can be considered working in zone A, however, for $\dot{\varepsilon} \leq 10^{-5}$ (1/s), creep  comes into scene, leaving a portion of the $\sigma_x$~=~0.5~mm conical target working under a combination of low stresses at high temperatures that may not be safe for the expected lifetime of the target, set to 155.4 days/year\footnote{The target lifetime is estimated assuming an operation cycle of 185 days per year \cite{FCC-ee_CDR} with a duty factor of 0.84 \cite{Ogur2018}}.

%Transient results
\subsection{Transient-state analysis}
\begin{figure}[b!]
    \centering
    \includegraphics[width=0.5\textwidth]{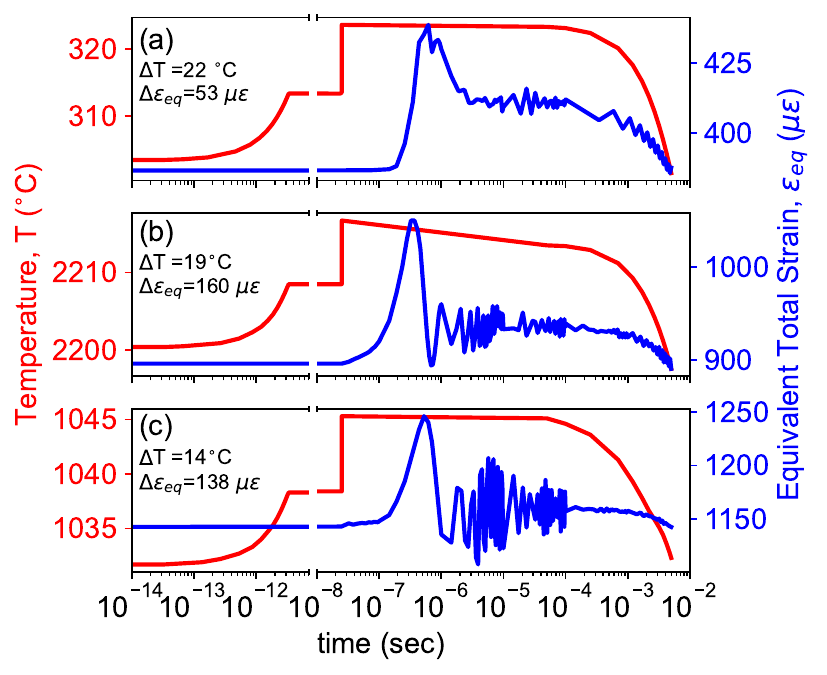} 
    \caption{\label{fig:Transient_TempStress} Transient-state for a single e$^-$ beam train impact after reaching steady-state: temperature at P1 (left-axis) and equivalent total strain at P2 (right-axis) vs time for the (a) baseline, (b) $\sigma_x=$~0.5~mm conical and (c) $\sigma_x=$~1~mm conical targets. Note: the plot is discontinuous for times \mbox{$t \in~$(7$\times10^{-12}$, 1$\times10^{-8}$) s}.}
\end{figure}

The transient behavior of the targets is summarized in Fig.~\ref{fig:Transient_TempStress}. Here, the effect of a primary e$^-$ train hitting the device is represented just after the steady-state regime is reached. For this analysis, the e$^-$ beam structure consists of one train containing two consecutive bunches with a duration of 3.33 ps and 25~ns apart~\cite{FCC-MTR2024}. The beam impact induces a staggered increment in temperature at P1, which generates stresses in P2, with a certain delay. As the targets are subjected to a multiaxial stress-strain state, the equivalent total strains $\varepsilon_{eq}$ are reported, instead of the equivalent stresses used in previous sections of the article. We observe a rise in temperature in two steps, according to the two-bunch time structure. Note that, as no thermal variation is registered along the 25~ns separation time, the target behaves as an adiabatic system. Thus, for representation purposes, the plots are discontinuous in \mbox{$t \in~$(7$\times10^{-12}$, 1$\times10^{-8}$) s}. Once the maximum temperature is reached, the dynamic response due to the rapid change of temperature is observed in the form of a peak in strains registered in the $\mu$s range, followed by an oscillatory return to the steady-state values during the absence of the beam, allowing the heat to be transferred to the surroundings of the target. This cyclic behavior occurs every 5~ms, when the following e$^-$ beam train impact is triggered. Depending on the target geometry, the temperature variation ranges $\Delta T \in [14,22]~^{\circ}$C in P1, resulting in total strains between $\Delta\varepsilon_{eq} \in [53,160]~\mu\varepsilon$ in P2. Due to the high repetition rate of the primary beam (200 Hz), there is not enough time for the cooling system to dissipate the internal heat generated in the target, and the fluctuations in temperature above the steady-state values are small. Consequently, the dynamic response in terms of strains is dissipated internally, allowing to return to the steady-state values without exceeding the elastic regime.

%\vfill
\subsection{Thermal fatigue assessment}
%Fatigue assessment
From the fatigue point of view, under the assumption that the target will be replaced every year, the required lifetime of the device is equivalent to 2.69$\times$10$^9$ thermal pulses\footnote{The total number of cycles are obtained by multiplying the target lifetime (days/year) by the electron bunch repetition rate, in this case 200 Hz.} induced by the primary e$^-$ beam hitting the tungsten part. This requirement is well beyond the 10$^6$ cycles criteria adopted in mechanical design for a part to be considered as infinite lifetime \cite{Shigley2011}. 

%Universal slope method (fatigue)
For the fatigue assessment, the selected methodology was the Universal Slope method, proposed by Manson \cite{Manson1965, Manson1966}, and implemented during the design of the JET and ITER divertors made of tungsten \cite{Mertens2013, Hirai2016}, respectively. This strain-life method is an empirical expression that relates the total strain range $\Delta\varepsilon_t$ to the number of cycles to failure $N_f$ to estimate the lifetime of a specimen in the form of the equation below:

\begin{eqnarray}
\label{eq:Universal_slope}
    \Delta\varepsilon_t &=&\Delta\varepsilon_p + \Delta\varepsilon_e \nonumber \\
    &=&D^{0.6}N_f^{-0.6}+3.5\frac{\sigma_u}{E}N_f^{-0.12}
\end{eqnarray}

\begin{figure}[ht!]
    \centering
    \includegraphics[width=0.49\textwidth]{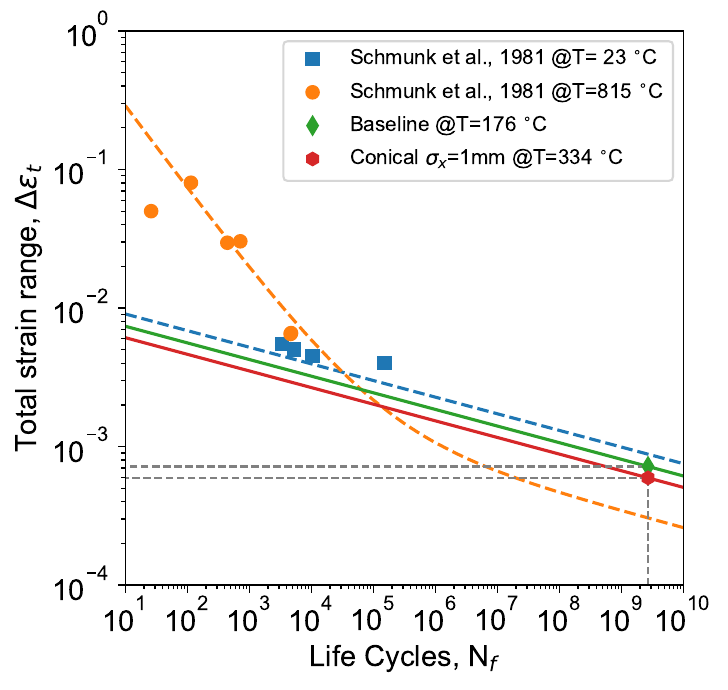} 
    \caption{\label{fig:US_tungsten} Universal slope curve for tungsten \cite{Schmunk1981} used to perform the thermal fatigue assessment for the FCC-ee e$^{+}$ source targets. }
\end{figure}

Where $\Delta\varepsilon_t$ is composed by the contribution of the plastic $\Delta\varepsilon_p$ and elastic $\Delta\varepsilon_e$ strains ranges. While $\Delta\varepsilon_p$ captures the behavior in the low cycle fatigue regime ($N_f\leq 10^3$) and depends on the material ductility $D$, $\Delta\varepsilon_e$ predicts the life in the high cycle fatigue regime ($N_f>10^3$) using only the ratio between the ultimate strength $\sigma_u$ with the Young's modulus $E$, where all material properties are temperature dependent. Fig.~\ref{fig:US_tungsten} contains the Universal slope curves for tungsten using the material values summarized in Table \ref{tab:Material_properties_for_fatigue}. The experimental data obtained by Schmunk et al., \cite{Schmunk1981} are used to show the validity of the method and how it is able to capture the behavior of tungsten below and above the ductile-to-brittle transition temperature (DBTT). For our analysis, the DBTT was set to 400 $^{\circ}$C, based on the behavior of tungsten at high strain-rate loading conditions reported in \cite{Scapin2019}.

%Fatigue assesment (results)
In Fig.~\ref{fig:US_tungsten}, using as an entry the expected lifetime of the target, and intersecting the line with the respective universal slopes for the baseline and conical target (green and red curves, respectively), the obtained total strain range is \mbox{$\Delta\varepsilon_t\in$ [595,720] $\mu\varepsilon$}. As $\Delta\varepsilon_{eq} < \Delta\varepsilon_t$, it is expected that any of the targets will not fail under thermal fatigue.

\subsection{Creep analysis}
\begin{figure}[t!]
    \centering
    \includegraphics[width=0.49\textwidth]{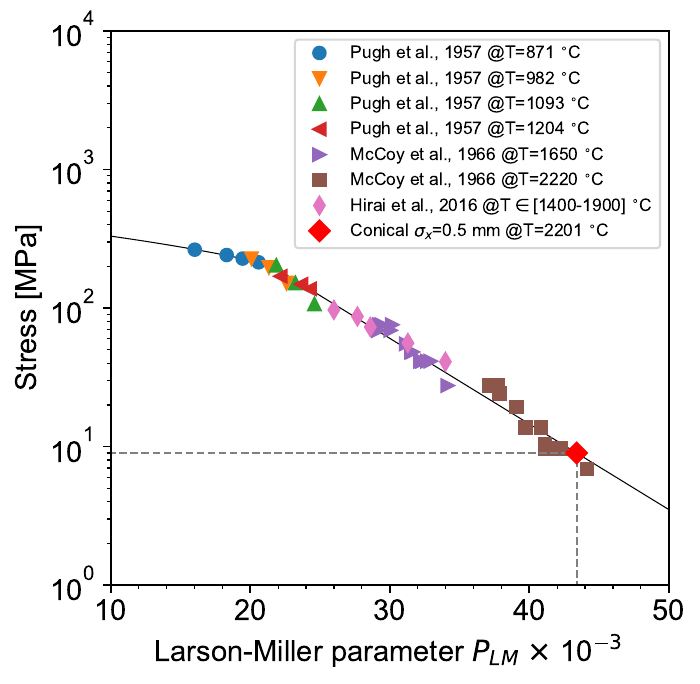} 
   \caption{\label{fig:LM_plot} Larson–Miller parameter ($P_{LM}$) vs. stress plot for tungsten obtained with the experimental data from \cite{Hirai2016, Pugh1957, McCoy1966}. }
\end{figure}

%Creep assesment (Larson-Miller parameter)
The Larson-Miller relation \cite{LarsonMiller1952} is used to estimate the impact of creep for the $\sigma_x$~=~0.5~mm conical target. It is expressed by the following equation:
\begin{equation}
\label{eq:LM_equation}
    P_{LM} = T(C + log~t_r)
\end{equation}

where $t_r$ is the time to rupture (hours), $T$ is the absolute temperature (K), and $C$ is a material constant equal to 15 for tungsten \cite{Hirai2016}. Fig.~\ref{fig:LM_plot} shows the Larson-Miller parameter ($P_{LM}$) vs stress plot  for tungsten obtained from the literature for temperatures in the range \mbox{$T\in [871,2220]~^{\circ}$C}. Based on the given curve, the $\sigma_x$~=~0.5~mm conical target can potentially fail around 14 days after reaching the steady state temperature and stresses at P1. Therefore, this e$^{+}$ target design is not suitable for the e$^-$ beam parameters of FCC-ee.

\subsection{Additional remarks}
%Radiation comment
Regarding the inclusion of radiation as a heat transfer mechanism, although the expected high temperatures in the protruding bodies were confirmed, its contribution respect to the total heat dissipation is negligible - being only 62 and 18 W for the $\sigma_x$~= 0.5 mm and $\sigma_x$~= 1 mm conical targets, respectively -  even when considering them  as perfect black bodies. For both configurations, the limiting factor corresponds to the small surface of the protruding geometries.

%Model limitation
As a limitation of the present study, the numerical models analyzed in this section assume that the material properties are not affected by the extreme ambient conditions. However, a preliminary study estimated the radiation damage, expressed in displacement per atom per year (DPA/year), of 1 DPA/year for both the baseline and conical target $\sigma_x$~= 1 mm. On the other hand, due to the lower beam size, the expected radiation damage  for the conical target $\sigma_x$~= 0.5 mm is  2 DPA/year.  These values are indicative and a proper experimental test campaign is needed to confirm the effect on the material properties.

\section{Implementation in the P\textsuperscript{3} experiment}
\label{sec:Implementation}

\begin{figure}[t!]
    \centering
    \includegraphics[width=0.5\textwidth]{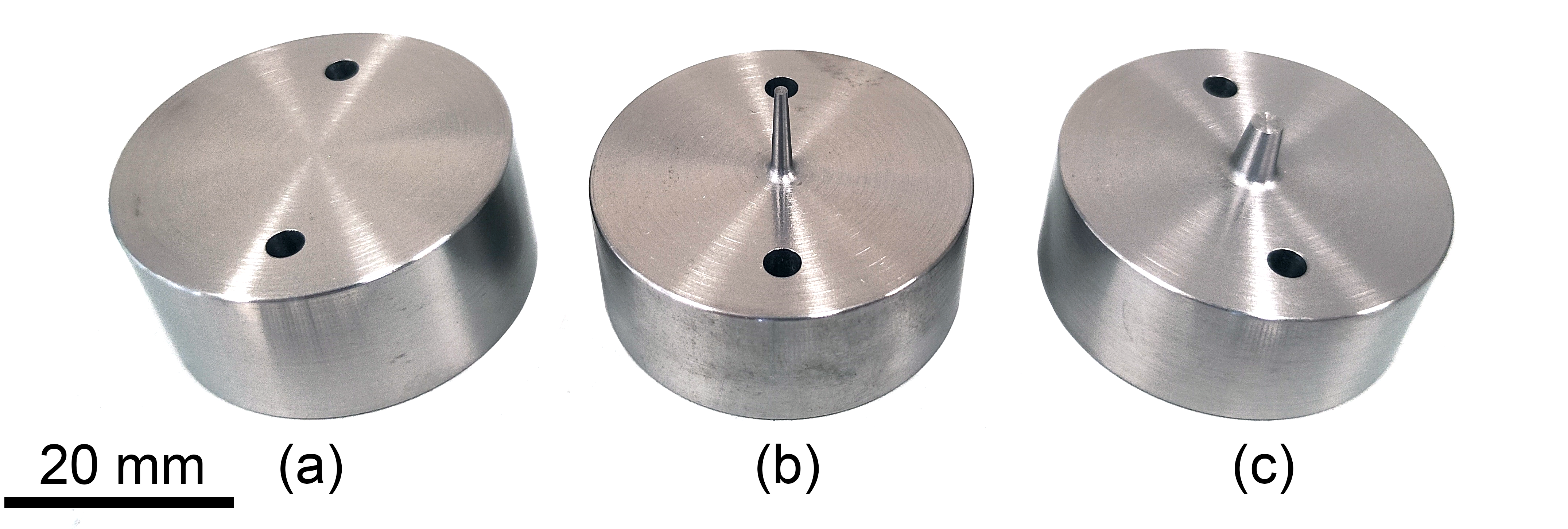} 
    \caption{\label{fig:P3_targets} Tungsten target prototypes for P$^3$: models without shielding and no cooling. Baseline cylindrical geometry (a), conical $\sigma_x$~=~0.5~mm (b) and conical $\sigma_x$~=~1~mm (c). The outside diameter is 37~mm and each target length $\ell_T$ is 17.5, 25 and 21 mm, respectively.}
\end{figure}

In order to test experimentally the e$^{+}$ yield enhancement from the conical targets and compare its performance with respect to the baseline cylindrical geometry, a target supporting device was developed as a part of the P$^3$ experiment. Following  the design specifications, the resulting device allows an easy installation, positioning and replacement of different targets \cite{Mena2024}. While the baseline target with embedded cooling pipes introduced in section \ref{sec:Thermo-Mechanical_Aspects} is still under a development phase, a series of tungsten targets without cooling pipes were manufactured as shown in Fig.~\ref{fig:P3_targets}, and they are ready to be tested. A future publication will explain the mechanical design of the P$^3$ target supporting device at a detailed level.

\section{Conclusion and Outlook}
\label{sec:VIIConc}

Our study shows that conical-shaped targets can significantly increase the e$^{+}$ yield of FCC-ee and its test facility P\textsuperscript{3}, using e$^+$ are based on a pair-production-driven scheme and will use a novel, highly efficient capture system based on an HTS solenoid. Furthermore, the increase in yield does not reduce significantly the beam quality in terms of transverse emittance or energy spread. As a result, conical targets not only boost the total e$^{+}$ production at the target but also improve the e$^{+}$ yield delivered to the DR, according to end-to-end simulations. Two conical target layouts were designed, considering two beam size options: $\sigma_x=$~0.5~mm and $\sigma_x=$~1~mm. The yield improvement provided by the $\sigma_x=$~1~mm option is significant: 60\% at the target and 16\% at the DR. More remarkably, the $\sigma_x=$~0.5~mm one could nearly duplicate the e$^{+}$ production at the target and increase the yield at the DR by around 60\%. 

Moreover, both conical targets were compared with respect to the FCC-ee baseline target design in terms of thermo-mechanical behavior. The $\sigma_x$=~1~mm conical target can withstand the current e$^-$ beam parameters for FCC-ee, although reaching significantly higher temperatures and thermal stresses—about three times greater than the baseline. However, the $\sigma_x$=~0.5~mm conical target is unsuitable for these conditions, with creep failure being the primary limiting factor. To address this limitation, further design upgrades are being investigated. First, a new baseline for the FCC-ee injector design was recently approved, based on a 2.86~GeV primary e$^-$ beam, a repetition rate of 100~Hz and four bunches per train~\cite{Craievich:FCC24}. Another promising solution is the addition of a thin tungsten crystal layer to the upstream surface of the target. According to parallel studies for FCC-ee, this approach could nearly halve the total power deposition within the target without compromising the e$^{+}$ yield~\cite{Alharthi:Channeling2024}. This enhancement would improve the thermo-mechanical resilience of the conical target, making it a viable option for high-power beam conditions. However, future work includes testing both conical targets in the P\textsuperscript{3} experiment, aiming at a proof-of-principle demonstration of the yield enhancement.

\appendix
%%%%%%%%%%%%%%%%%%%%%%%%%%
\section*{Appendix}
%%%%%%%%%%%%%%%%%%%%%%%%%%
% Change the section numbering style
\renewcommand{\thesection}{\Alph{section}} % Sections now use A, B, C...
\setcounter{section}{0} % Reset section counter

% Redefine numbering for figures and tables
\renewcommand{\thefigure}{A\arabic{figure}}
\renewcommand{\thetable}{A\arabic{table}}
% Reset counters
\setcounter{figure}{0}
\setcounter{table}{0}

%Material properties
\section{Material properties}
\label{sec:Material_properties}

The material properties as a function of temperature required for the thermo-mechanical simulation presented in Section \ref{sec:Thermo-Mechanical_Aspects} were the density $\rho$, specific heat capacity $C_p$ and thermal conductivity $k$, for the thermal step. In addition, for the mechanical step, the Young’s modulus $E$, the Poisson’s coefficient $\nu$, the yield stress $\sigma_y$, and the thermal expansion coefficient $\alpha$ were used. All data were taken from the library Material Properties Data Base (MPDB) \cite{MPDB2024}.

\begin{figure}[t!]
    \centering
    \includegraphics[width=0.5\textwidth]{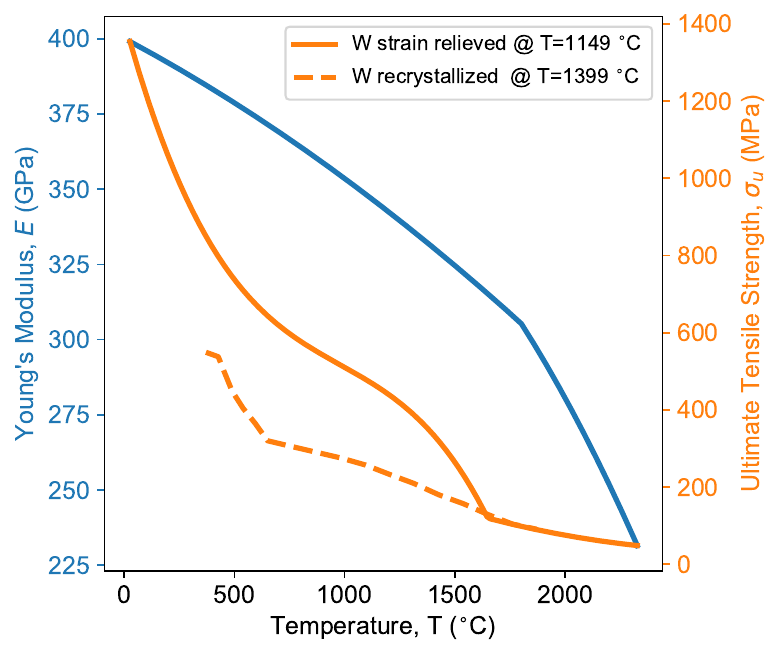} 
    \caption{\label{fig:E_vs_stu_plot} Young's Modulus $E$ and ultimate tensile strength $\sigma_u$ as a function of temperature $T$ for tungsten \cite{MPDB2024, Dotson1964}. }
\end{figure}

Based on Eq.~\ref{eq:Universal_slope}, for the elastic contribution of the strain range $\Delta\varepsilon_e$, the ratio $\frac{\sigma_u}{E}$ was needed. Fig.~\ref{fig:E_vs_stu_plot} represents each material property in the temperature range of interest for the target design \mbox{$T\in[27,2320]~^{\circ}C$}, where the ultimate strength $\sigma_u$ for recrystallized tungsten was included \cite{Dotson1964} (see dashed line) to show the reduction in material properties respect to the strain relieved condition used in the numerical model.

The material properties used to generate the Universal Slopes for tungsten presented in Fig.~\ref{fig:US_tungsten} are summarized on Table \ref{tab:Material_properties_for_fatigue}. Here, RA corresponds to the reduction in area ($\%$) and the ductility $D$ is obtained using the relation $D=ln\frac{1}{1-RA}$. For all cases were \(T \leq 400^{\circ}C\), tungsten was considered as brittle and therefore, $D$ was set to zero.

\begin{table}[t!]
\caption{Material properties used to generate the Universal Slopes for tungsten presented in Fig.~\ref{fig:US_tungsten}.}
\label{tab:Material_properties_for_fatigue}
\begin{center}
\begin{tabular}{ ccccc } 
\textbf{T}& \textbf{RA }& \textbf{$\sigma_u$} & \textbf{E}& \textbf{Ref.}\\
\textbf{[$^{\circ}$C]}& \textbf{[\%]}& \textbf{[MPa]} & \textbf{[GPa]}& \\
 \hline
 23  & - &1362.1&399.2  & \cite{Dotson1964} \\ 
 176 & - &1093.5&393.0  & \cite{Dotson1964}\\ 
 306 & - &920.5 &387.5  & \cite{Dotson1964}\\ 
 334 & - &889.0 &386.2  &\cite{Dotson1964}\\
 815 &71 &426.5 &363.3  & \cite{Schmunk1981, Dotson1964}\\
 \hline
\end{tabular}
\end{center}
\end{table}

%%%%%%%%%%%%%%%%%%%%%
\section*{Acknowledgemets}
%%%%%%%%%%%%%%%%%%%%%
This work was done under the auspices of CHART (Swiss Accelerator Research and Technology) and the Future Circular Collider Innovation Study (FCCIS). This project has received funding from the European Union’s Horizon 2020 research and innovation programme under grant agreement No 951754.

%%% References %%%

\end{document}